\documentclass[aps,prd,reprint,groupedaddress,showpacs]{revtex4-1}

\usepackage{graphicx}
\usepackage{dcolumn}
\usepackage{bm}

\newcommand{\pom}{\tt I\! P}

\begin{document}

\title{Deciphering  charmoniumlike exotic states in photon - hadron interactions at  RHIC and LHC energies}

\author{V. P. Gon\c calves}

\email[]{barros@ufpel.edu.br}

\affiliation{Instituto de F\'{\i}sica e Matem\'atica, Universidade Federal de
Pelotas\\
Caixa Postal 354, CEP 96010-090, Pelotas, RS, Brazil}

\author{M. L. L. da Silva}

\email[]{mario.silva@ufpel.edu.br}

\affiliation{Instituto de F\'{\i}sica e Matem\'atica, Universidade Federal de
Pelotas\\
Caixa Postal 354, CEP 96010-090, Pelotas, RS, Brazil}

\begin{abstract}
In this paper we propose the study of the photoproduction of charmoniumlike states in $pp$ collisions at RHIC and LHC energies and estimate the rapidity distribution and total cross sections for the 
production of the exotic hadrons $Y(3940)$, $X(3915)$, $Z(4430)$ and $Z_c(3900)$. We demonstrate that the  experimental analysis of this process  can provide complementary and independent checks on these states, and help to understand their underlying nature.
\end{abstract}


\maketitle

\section{Introduction} 
 In the past eleven  years a series of charmoniumlike states $X, Y, Z$ has been announced at various experimental facilities (For recent reviews see, {\it e.g.}, Refs. \cite{bambrila,pr_navarra,navarra_mpla}). These exotic mesons are a class of hadrons that decay to final states that contain a heavy quark and a heavy antiquark but cannot be easily accommodated in the remaining unfilled states in the $c\bar{c}$ level scheme.  One of the most interesting exotic states are the charged states, {\it e.g.} $Z(4430)$ and $Z_c(3900)$, that clearly have a more complex structure than $c\bar{c}$ pair, being natural candidates for molecular or tetraquarks states.

So far, most of the available experimental and theoretical investigations of the charmoniumlike exotic candidates have been focused on the spectrum and decays or the production in $e^+ e^-$ collisions. In order to decipher the nature of these states, more accurate data and new processes involving these states would be equally important. The search of $X,Y,Z$ charmoniumlike states by other production processes  will not only confirm these observed states, but also will be useful to study their underlying structures. In the last years, the study of the production of exotic hadrons in proton - proton  \cite{exoticpp}, heavy ion \cite{exotic_heavy} and photon - hadron \cite{z4430,y3940,zc3900,x3915} collisions was proposed by several authors.  In particular, the results obtained in Refs.  \cite{z4430,zc3900} for the photoproduction of  charged charmoniumlike states $Z(4430)$ and $Z_c(3900)$ indicate a large enhancement of the cross section near the threshold, which could be considered a signature for the existence of these states in $\gamma p$ collisions. Unfortunately, 
as the DESY - HERA $ep$ collider stopped its operations in 2007, the experimental analysis of these states in photon - hadron collisions was not possible and remains an open question.

In recent years we  proposed the analysis of photon  induced interactions in hadronic  collisions  as an alternative way to study the QCD dynamics at high energies
\cite{vicmag_upcs}. 
The basic idea  is that in this interactions the total cross section for a given process can be factorized in
terms of the equivalent flux of photons into the hadron projectile and the photon-photon or photon-target production cross section. The main advantages of using hadron - hadron collisions for
studying photon induced interactions are the high equivalent photon
energies and luminosities that can be achieved at existing  accelerators (For a review see Ref. \cite{upcs}). In Ref. \cite{gama_prc} we estimated the production of some exotic states in $\gamma \gamma$ interactions  considering $pp$ collisions at LHC energies and demonstrated that the experimental analysis of these states is, in principle, feasible. A similar conclusion was obtained in Ref. \cite{gama_bert} for $PbPb$ collisions. 
In this paper we will demonstrate that the study of $\gamma p$ interactions
at the RHIC and LHC can also provide complementary and independent checks on the properties of the  exotic states, and help to understand their underlying nature.
In this paper,  we obtain, for the first time, an estimate of the photoproduction of the charmoniumlike states 
$Y(3940)$, $X(3915)$, $Z(4430)$ and $Z_c(3900)$ in proton - proton collisions at the RHIC and LHC. 
As we will show in the following, the prospect to observe these states  is rather promising and thus an  experimental analysis is strongly motivated.

This paper is organized as follows. In the next section we present a brief review of photon - hadron interactions in hadron - hadron collisions. In Section \ref{photo_exotic} we discuss the photoproduction of exotic charmoniumlike states and present the assumptions of the formalism used in our calculations. In Section \ref{resultados} we present our predictions for the 
rapidity distribution and total cross sections for the 
photoproduction of the exotic hadrons $Y(3940)$, $X(3915)$, $Z(4430)$ and $Z_c(3900)$ in $pp$ collisions at RHIC and LHC energies. Finally, in Section \ref{conc} we present a summary of our main conclusions.

\section{Photon - hadron interactions in $pp$ collisions}  
Lets consider a hadron-hadron interaction at large impact parameter ($b > R_{h_1} + R_{h_2}$) and at ultra relativistic energies. In this regime we expect the dominance of the electromagnetic interaction.
In  heavy ion colliders, the heavy nuclei give rise to strong electromagnetic fields due to the coherent action of all protons in the nucleus, which can interact with each other. In a similar way, it also occurs when considering ultra relativistic  protons in $pp(\bar{p})$ colliders.
The photon stemming from the electromagnetic field
of one of the two colliding hadrons can interact with one photon of
the other hadron ($\gamma \gamma$ process) or can interact directly with the other hadron ($\gamma h$
process). For photon - hadron interactions the total
cross section for a given process can be factorized in terms of the equivalent flux of photons of the hadron projectile and  the  photon-target production cross section \cite{upcs}. 
In the particular case of the photoproducion of a charmoniumlike state $H_c$  in a  hadron-hadron collision, the total cross section is  given by
\begin{eqnarray}
\sigma (h_1 h_2 \rightarrow p \otimes H_c h_3 ) =  \sum_{i=1,2} \int dY \frac{d\sigma_i}{dY}\,,
\label{sighh}
\end{eqnarray}
where  $h_1 = h_2 = p$, $\otimes$ represents a rapidity gap in the final state and $h_3 = p$ or $n$ depending if the interaction is mediated by a neutral ($\omega$) or a charged ($\pi^+$) particle, respectively.  Moreover,  ${d\sigma_i}/{dY}$ is the rapidity distribution for the photon-target interaction induced by the hadron $h_i$ ($i =1,2$), which can be expressed as 
\begin{eqnarray}
\frac{d\sigma_i}{dY} = \omega \frac{dN_{\gamma/h_i}}{d\omega}\,\sigma_{\gamma h_j \rightarrow H_c h_3} (W_{\gamma h_j}^2) \,\,\,\,\,\,(i\neq j)\,.
\label{rapdis}
\end{eqnarray}
where $\omega$ is the photon energy,   $\frac{dN_{\gamma/h}}{d\omega}$ is the equivalent photon flux, $W_{\gamma h}^2=2\,\omega \sqrt{s_{\mathrm{NN}}}$  and ${s_{\mathrm{NN}}}$ are  the  c.m.s energy squared of the
photon - hadron and hadron-hadron system, respectively. 
In what follows we assume that the  photon spectrum of a relativistic proton is given by  \cite{Dress},
\begin{widetext}
\begin{eqnarray}
\frac{dN_{\gamma/p}(\omega)}{d\omega} =  \frac{\alpha_{\mathrm{em}}}{2 \pi\, \omega} \left[ 1 + \left(1 -
\frac{2\,\omega}{\sqrt{s_{NN}}}\right)^2 \right] .
\left( \ln{\Omega} - \frac{11}{6} + \frac{3}{\Omega}  - \frac{3}{2 \,\Omega^2} + \frac{1}{3 \,\Omega^3} \right) \,,
\label{eq:photon_spectrum}
\end{eqnarray}
\end{widetext}
where $\Omega = 1 + [\,(0.71 \,\mathrm{GeV}^2)/Q_{\mathrm{min}}^2\,]$, $Q_{\mathrm{min}}^2= \omega^2/[\,\gamma_L^2 \,(1-2\,\omega /\sqrt{s_{NN}})\,] \approx (\omega/
\gamma_L)^2$ and $\gamma_L$ is the Lorentz factor.
The experimental separation for such events is relatively easy, as photon emission is coherent over the  hadron and the photon is colorless we expect the events to be characterized by intact recoiled hadron (tagged hadron) and the presence of a  rapidity gap (For a detailed discussion see \cite{upcs}). Moreover, the detection of the neutron in the final state can be useful to separate the events associated to the production of a charged charmoniumlike state. Finally, it is important to emphasize that as  the   photon spectrum decreases  with energy approximately like $1/\omega$,   the main contribution for the total cross section, Eq. (\ref{sighh}), comes from the small $\omega$ behaviour of the photon spectrum where we are probing the photon - hadron cross section at low values of the center-of-mass energy.

\begin{figure}[t]
\begin{tabular}{cc}
\includegraphics[scale=0.28]{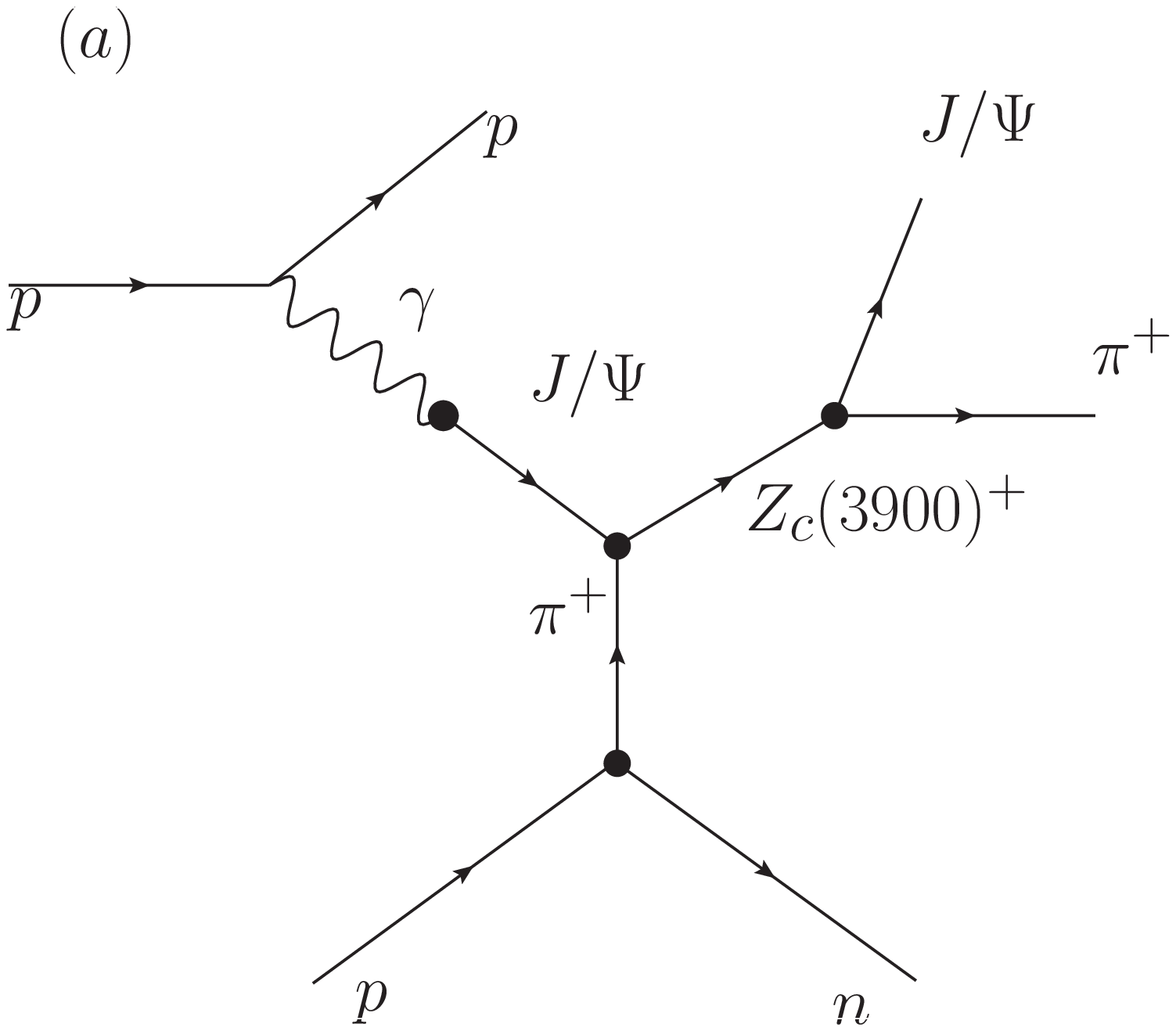} &
\includegraphics[scale=0.25]{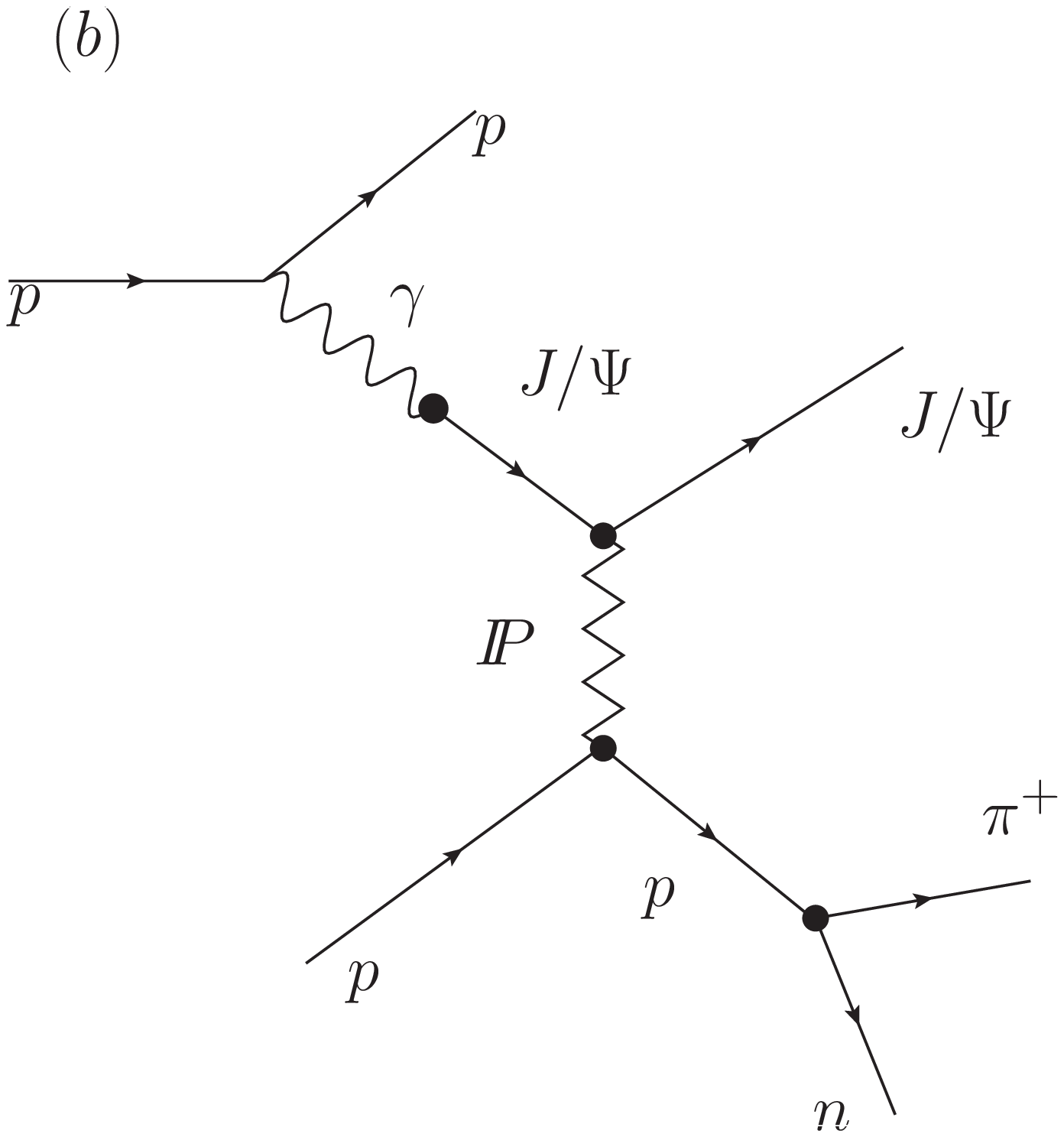}
\end{tabular}
\caption{(a) The $pp \rightarrow p Z_c^+ n \rightarrow p  J/\Psi \pi^+ n$ process through the $\pi^+$ exchange. (b) The $pp \rightarrow p  J/\Psi \pi^+ n$ process through the Pomeron exchange. }
\label{diagrama}
\end{figure}

\section{Photoproduction of exotic charmoniumlike states}
\label{photo_exotic} 

 The main input in our calculations is the photon - hadron cross section for the production of the  of exotic charmoniumlike state $\sigma_{\gamma p \rightarrow H_c}$.
Currently the unique available predictions in the literature \cite{z4430,y3940,zc3900,x3915} were obtained considering an effective Lagrangian approach combined with the vector meson dominance (VMD) assumption \cite{vdm_3}. In this approach the interaction is described in terms of a meson exchange, which is neutral/charged  for the production of neutral/charged exotic charmoniumlike states. In what follows we  briefly review the main assumptions of the approach proposed in Refs.  \cite{z4430,y3940,zc3900,x3915}, with particular emphasis in the production of the $Z_c(3900)^+$. The results for the other exotic states are obtained in a similar way.

Considering the approach proposed in Ref. \cite{zc3900} the photoproduction of the $Z_c(3900)^+$ in $pp$ collisions is described by the diagram presented in Fig. \ref{diagrama}(a), where we also represent  the decay of the exotic meson in a $J/\Psi + \pi^+$ final state, which we assume to be the dominant channel. The basic idea is that the  photon stemming from the electromagnetic field of one of the incident protons  fluctuates into a $J/\Psi$ which interacts with the other proton through the $\pi^+$ exchange producing a neutron $n$ and a $Z_c(3900)^+$ state which decays in the  $J/\Psi +  \pi^+$ system.
Assuming the VMD model to describe the coupling between the intermediate vector meson and the photon, an effective Lagrangian to describe the coupling between the pion and the nucleons and another describing the $Z_c J/\Psi \pi$ coupling,   the squared amplitude for the process $\gamma p \rightarrow  Z_c^+ n$ can be expressed by (See Ref. \cite{zc3900} for details)
\begin{eqnarray}
|{\cal{M}}|^2 & = & {\cal{C}} \frac{-q^2(q^2 - M_{Z_c}^2)^2}{(q^2 - m_{\pi}^2)^2} F_{\pi NN}^2 F_{Z_cJ/\Psi \pi}^2
\end{eqnarray}
where $q$ is the four momentum of the pion exchanged and
\begin{eqnarray}
{\cal{C}} = \left(\sqrt{2} g_{\pi NN} \frac{g_{Z_cJ/\Psi \pi}}{M_{Z_c}} \frac{e}{f_{J/\Psi}}\right)^2
\end{eqnarray}
with the  couplings $g_{\pi NN}$, ${e}/{f_{J/\Psi}}$ and $g_{Z_cJ/\Psi \pi}$ being respectively determined  assuming that $g_{\pi NN}^2/4 \pi = 14$, by the decay $J/\Psi \rightarrow e^+ e^-$ and by the decay width of $Z_c \rightarrow J/\Psi \pi$. In what follows we assume $\Gamma[Z^+ \rightarrow J/\Psi \pi^+] = 29$ MeV as calculated in \cite{maiani}. Moreover, the form factors $F_{\pi NN}$ and $F_{Z_cJ/\Psi \pi}$  are given by
\begin{eqnarray}
 F_{\pi NN}  =   \left(\frac{\Lambda_{\pi}^2 - m_{\pi}^2}{\Lambda_{\pi}^2 - q^2}\right) \,, \,\,\,
 F_{Z_cJ/\Psi \pi}  =   \left(\frac{\Lambda_{Z_c}^2 - m_{\pi}^2}{\Lambda_{Z_c}^2 - q^2}\right)
\end{eqnarray}
where we assume   $\Lambda_{\pi} = 0.7$ GeV and $\Lambda_{Z_c} = M_{J/\Psi}$ for the cut-offs. The resulting $\gamma p$ cross section is strongly enhanced close to the threshold \cite{zc3900}. 

\begin{figure}[t]
\includegraphics[scale=0.35]{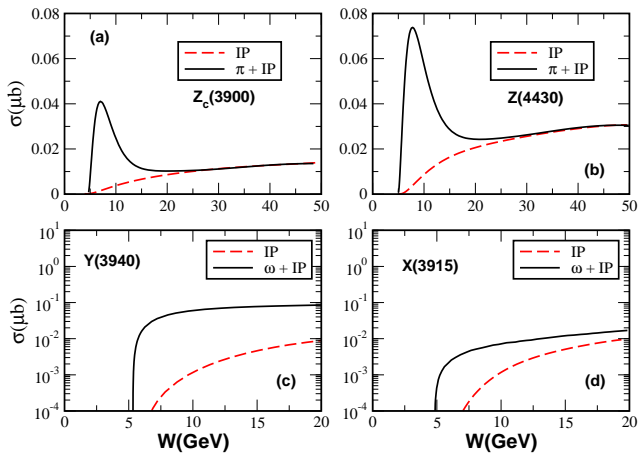} 
\caption{(Color online) Energy dependence of the photoproduction cross sections. The signal (Meson + $\pom$) and background ($\pom$) contributions are presented separately.  }
\label{gamap}
\end{figure}

This formalism can be directly extended to take into account the $Z_c$ decay in the $J/\Psi + \pi$ final state. However, as this final state can also be produced in a $\gamma p$ interaction through the Pomeron exchange, we should include this background in our calculations. In  Fig. \ref{diagrama}(b) we present a typical diagram for the background contribution. Following Ref. \cite{ln} we assume that the Pomeron behaves like an isoscalar photon with $C=+1$. The corresponding amplitude has a Regge-like energy dependence $\propto s^{\alpha_{\pom}(t) - 1}$, where   $\alpha_{\pom}(t) = 1 + \Delta + \alpha^{\prime}t$ is the Pomeron trajectory, $\Delta$ is the Pomeron intercept, $\alpha^{\prime} = 0.25$ GeV$^{-2}$ and $t$ is the exchanged Pomeron momentum squared. In Fig. \ref{gamap}(a) we present our predictions for the energy dependence of the $\gamma p \rightarrow J/\Psi \pi n$ cross section considering the signal ($\pi + \pom$) and background ($\pom$) contributions. We obtain that  the resulting cross section for the $\gamma p \rightarrow J/\Psi \pi n$ process is dominated by the signal for $W_{\gamma p} \le 20$ GeV, in agreement with the results presented in Ref.  \cite{zc3900}.

\begin{figure}[t]
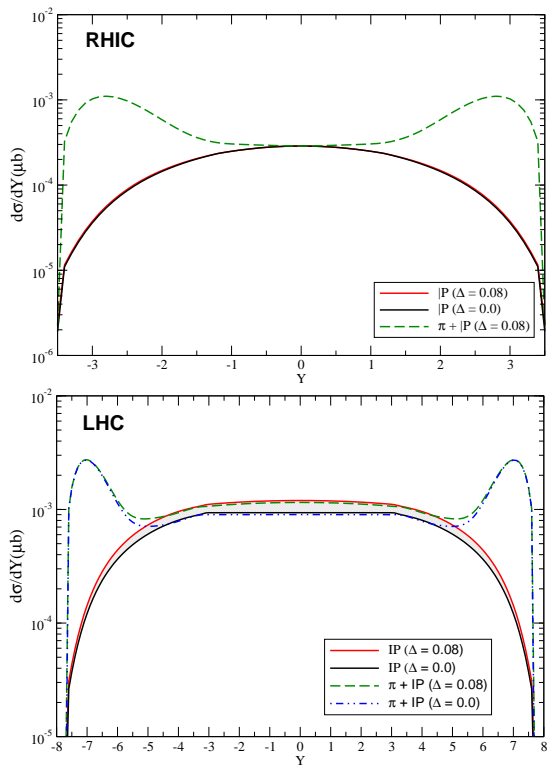

\includegraphics[scale=0.3]{Zc3900_rhic200.eps} \\
\includegraphics[scale=0.3]{Zc3900_lhc14.eps}
\caption{(Color online) Rapidity distribution for the photoproduction of a $J/\Psi + \pi$ final state in $pp$ collisions at RHIC ($\sqrt{s} = 0.2$ TeV) and LHC ($\sqrt{s} = 14$ TeV) energies.  The solid lines represent the background associated to the  Pomeron exchange for two values of the intercept $\Delta$.  The dashed lines represent the sum of the background with the signal associated to the $\gamma p \rightarrow Z_c(3900)^+ n \rightarrow J/\Psi \pi n$ interaction through $\pi$ exchange.}
\label{zc3900_14}
\end{figure}

\begin{figure}[ht]
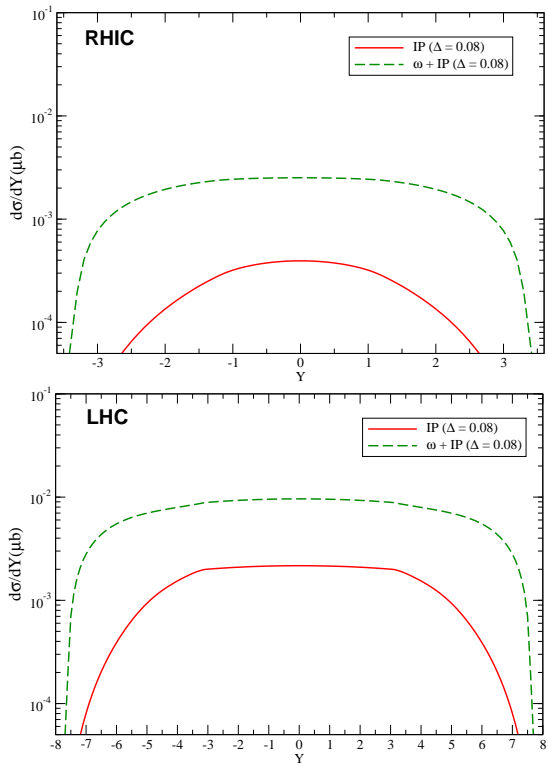

\includegraphics[scale=0.3]{Y3940_rhic200.eps} \\
\includegraphics[scale=0.3]{Y3940_lhc14.eps}
\caption{(Color online) Rapidity distribution for the photoproduction of a $J/\Psi + \omega$ final state in $pp$ collisions at RHIC ($\sqrt{s} = 0.2$ TeV) and LHC ($\sqrt{s} = 14$ TeV) energies.  The solid lines represent the background associated to the  Pomeron exchange ($\Delta$ = 0.08).  The dashed lines represent the sum of the background with the signal associated to the $\gamma p \rightarrow Y(3940)+ p \rightarrow J/\Psi \omega p$ interaction through $\omega$ exchange.}
\label{y3940}
\end{figure}

A similar analysis can also be performed for the photoproduction of the $Z(4430)^+$. The main differences are that we assume that   the coupling of the photon occurs with a $\Psi^{\prime}$ instead of the $J/\Psi$ and that  the dominant decay channel of the $Z(4430)^+$ is in a $\Psi^{\prime} + \pi^+$ system. Following Ref. \cite{z4430} 
 we also assume a $Z(4430)$ with $J^P = 1^-$ quantum numbers. However, we expect to obtain similar results for a state with  $J^P = 1^+$. As in the  $Z_c(3900)^+$ case, we obtain in Fig. \ref{gamap}(b) that the $\sigma (\gamma p \rightarrow \Psi^{\prime} \pi n)$ is dominated by the signal at small values of the $\gamma p$ center-of-mass energy, with a strong  enhancement close to the threshold, reproducing the results presented  in Ref. \cite{z4430} for $\Lambda_{Z \rightarrow \Psi^{\prime} \pi} = 45$ MeV and $g_{Z \Psi^{\prime} \pi} = 2.365$ GeV$^{-1}$. Finally, for the photoproduction of the neutral charmoniumlike states 
 $Y(3940)$  and  $X(3915)$ we assume that these states are  $0^{++}$ and $2^{++}$ states, that  
 the cross sections for the signals are described in terms of a $\omega$ exchange and that these states decay into a $J/\Psi + \omega$. It is  important to emphasize that distinctly from the production of charged states, for the case of neutral states,  the proton target remains intact in the final state. Consequently, the final state for the photoproduction of these states in $pp$ collisions is characterized by two intact protons. Our predictions for  $\sigma (\gamma p \rightarrow J/\Psi \pi p)$ are presented in Figs. \ref{gamap}(c) and (d)   for $\Gamma[Y(3940) \rightarrow J/\Psi \omega] = 34$ MeV and  $\Gamma[X(3915) \rightarrow J/\Psi \omega] = 5.10$ MeV and are similar to those obtained in Refs. \cite{y3940,x3915}. It is important to emphasize that for these states the total cross section do not have a resonancelike structure close to the threshold, with the signal being larger than  the background for  $W_{\gamma p} \le 50 \,(20)$ GeV  for the  $Y(3940)$ ($X(3915)$) photoproduction.

\section{Results}\label{resultados}
  Lets initially calculate the rapidity distribution and total cross section for the photoproduction of the charged charmoniumlike states $Z_c(3900)$ and $Z(4430)$. 
The distribution on rapidity $Y$ of the hadron $H_c$ of mass $M_{H_c}$ in the final state can be directly computed from Eq. (\ref{sighh}), by using its  relation with the photon energy $\omega$, i.e. $Y\propto \ln \, ( \omega/M_{H_c})$.  Explicitly, the rapidity distribution is written down as,
\begin{widetext}
\begin{eqnarray}
\frac{d\sigma \,\left[h_1 h_2 \rightarrow   p \otimes H_c + n \right]}{dY} = \left[\omega \frac{dN}{d\omega}|_{h_1}\,\sigma_{\gamma h_2 \rightarrow  H_c + n}\left(\omega \right)\right]_{\omega_L} + \left[\omega \frac{dN}{d\omega}|_{h_2}\,\sigma_{\gamma h_1 \rightarrow H_c + n}\left(\omega \right)\right]_{\omega_R}\,
\label{dsigdy2}
\end{eqnarray}
\end{widetext}
where  $\omega_L \, (\propto e^{-Y})$ and $\omega_R \, (\propto e^{Y})$ denote photons from the $h_1$ and $h_2$ hadrons, respectively.  As the photon fluxes, Eq. (\ref{eq:photon_spectrum}), have support at small values of $\omega$, decreasing exponentially at large $\omega$, the first term on the right-hand side of the Eq. (\ref{dsigdy2}) peaks at positive rapidities while the second term peaks at negative rapidities. Consequently, given the photon flux, the study of the rapidity distribution can be used to constrain  the photoproduction cross section for a given energy. Moreover, the total rapidity distributions for $pp$  collisions will be symmetric about midrapidity ($Y=0$).

In Fig. \ref{zc3900_14} we present our predictions for the rapidity distribution for the production of a $J/\Psi + \pi$ final state in $pp$ collisions at RHIC ($\sqrt{s} = 0.2$ TeV) and LHC ($\sqrt{s} = 14$ TeV) energies considering the diagrams presented in Figs. \ref{diagrama} (a) and (b).  The cross section for the background, associated to the Pomeron exchange, has been estimated considering two different values for the Pomeron intercept $\Delta = 0$ and 0.08, which determines its energy dependence. At RHIC energy these two predictions for the Pomeron contribution are almost identical, while at LHC they differ by almost 10 $\%$ at $Y = 0$. We predict that the signal dominates the rapidity distributions for large values of the rapidity $|Y|$, where we are probing the $\gamma p$ cross section at small values of the c.m. energy. The resonancelike structure  present in the $\gamma p$ cross section, which is a signature of the $Z_c^+$ production, can be directly probed at $|Y| \geq 2.0 \, (5.5)$ at RHIC (LHC) energy. A similar behaviour is predicted in the rapidity distribution for  the photoproduction of the $Z(4430)^+$.

Our predictions for the photoproduction of the neutral charmoniumlike state $Y(3940)$ decaying into a $J/\Psi + \omega$ state are presented in Fig. 
\ref{y3940}. In this case we predict that the signal is larger than the background  at all values of rapidity.     In particular, we predict that the signal is larger by a factor $\ge 5$ at $Y = 0$ . A similar conclusion is reached for the $X(3915)$.

In Table \ref{tab1} we present our predictions for the total cross sections for the different final states discussed above. In particular, we present  the predictions for the background contributions as well as for the signals which take into account the production of  different exotic charmoniumlike states. We predict that the cross sections are of $\cal{O}$(nb) for RHIC and LHC energies. It is important to emphasize that similar numbers are obtained for the Tevatron energy. For the charged states the signal is larger than the background by a factor $\ge 1.6$, being about three for the RHIC energy. In order to estimate the number of events in $pp$ collisions at LHC, we assume the  design luminosity ${\cal L} = 10^7$ mb$^{-1}$ s$^{-1}$. Consequently, we predict  ${\cal{O}}(10)$ events/second for these processes.

\section{Conclusions} \label{conc}
 During the last decade a rich spectroscopy of charmoniumlike states that do not fill into the remaining unassigned levels for $c\bar{c}$ charmonium states has emerged. However, several questions still remain to be answered. In this paper we have proposed, for the first time, the study of the photoproduction of exotic charmoniumlike states in $pp$ collisions at RHIC and LHC energies. Considering an effective Lagrangian description for the photon - hadron interaction we have derived an estimate of the production rates for these particles. Our results show that the resulting cross sections for the  $Y(3940)$, $X(3915)$, $Z(4430)$ and $Z_c(3900)$ photoproduction at the RHIC energy are at 1 nb level, while at LHC are larger by roughly one order of magnitude. 
Taking into account the design luminosity at the LHC the number of events has been estimated. Our results indicate that these four exotic states could be copiously produced. Consequently, we believe that study of these states at RHIC and LHC is feasible and  will provide valuable information on hadron spectroscopy as well as hadron interactions.
Finally, measurements of these exotic states at RHIC and LHC will supplement the results obtained at $e^+e^-$ colliders, and thus allow to explore the nature of the states.

\begin{widetext}

\begin{table}[t]
\begin{center}
\begin{tabular} {|c|c|c|c|c|c|}
\hline
\hline
Reaction & Ressonance & Contribution & $\sigma$ [nb] ($\sqrt{s} = 0.2$ TeV) 
 & $\sigma$ [nb] ($\sqrt{s} = 7$ TeV) & $\sigma$ [nb] ($\sqrt{s} = 14$ TeV) \\
\hline
\hline
$\sigma(pp \rightarrow pJ/\Psi\pi n)$ & -- & $\pom$ & 1.15 & 8.18 -- 9.64 & 10.33 -- 12.65 \\
 &$Z_c(3900)$ & $\pom + \pi$ & 3.83 & 14.13 -- 15.52 & 16.89 -- 19.12 \\
\hline
$\sigma(pp \rightarrow p\Psi^{\prime}\pi n)$  & -- & $\pom$ & 2.60 & 18.15 -- 21.32 & 22.87 -- 27.93 \\
 &$Z(4430)$ & $\pom + \pi$ & 7.33 & 29.26 -- 32.41 & 35.21 -- 40.23 \\
 \hline
$\sigma(pp \rightarrow pJ/\Psi\omega p)$ & -- & $\pom$ & 0.84 -- 0.90 & 5.90 -- 7.75 & 7.42 -- 10.17 \\
& $X(3915)$  & $\pom + \omega$ & 1.88 -- 1.98 & 11.31 -- 14.53 & 14.08 -- 18.88 \\
\hline
$\sigma(pp \rightarrow pJ/\Psi\omega p)$ & -- & $\pom$ & 1.33 & 12.73 -- 15.35 & 16.35 -- 20.54 \\
& $Y(3940)$  & $\pom + \omega$ & 12.62 & 74.28 -- 85.93 & 92.58 -- 111.19 \\
\hline
\hline
\end{tabular}
\end{center}
\caption{Total cross sections for the photoproduction of different final states in $pp$ collisions at RHIC and LHC energies considering the sum of the signal associated to the photoproduction of an exotic charmoniumlike state, produced by a $\pi$ or $\omega$ exchange, and the background contribution associated to the Pomeron ($\pom$) exchange. For comparison the magnitude of the background contribution is presented separately.}
\label{tab1}
\end{table}

\end{widetext}

\section*{Acknowledgments} 
The authors are grateful to A. Martinez Torres, K. Kemchandani, F. S. Navarra and M. Nielsen for discussions.
This work has been supported by CNPq and FAPERGS, Brazil.

\end{document}